\documentclass[sigconf,screen]{acmart}

\usepackage{booktabs}
\usepackage{multirow}
\usepackage{longtable}
\usepackage{graphicx}
\usepackage{float}

\AtBeginDocument{%
  }

\setcopyright{none}
\renewcommand{\footnotetextcopyrightpermission}[1]{}
\begin{document}

\title{Wearing Trust: How Older Adults Calibrate Reliance on Health Wearables Through Bodily Experience and Everyday Use}

\author{Yibo Meng}
\affiliation{%
  \institution{Tsinghua University}
  \city{Beijing}
  \country{China}
}
\email{mengyb22@tsinghua.org.cn}

\author{Bingyi Liu}
\affiliation{%
  \institution{University of Michigan}
  \city{Ann Arbor}
  \country{USA}
}
\email{Bingyi.ux@gmail.com}

\author{ZhiMing Liu}
\affiliation{%
  \institution{University of Shanghai for Science and Technology}
  \city{Shanghai}
  \country{China}
}
\email{2120600527@st.usst.edu.cn}

\author{Ruiqi Chen}
\affiliation{%
  \department{Human Centered Design \& Engineering}
  \institution{University of Washington}
  \city{Seattle}
  \country{USA}
}
\email{ruiqich@uw.edu}

\renewcommand{\shortauthors}{Yibo Meng, Bingyi Liu, ZhiMing Liu, and Ruiqi Chen}

\begin{abstract}
Older adults increasingly use health wearables, yet often cannot inspect the properties that matter for reliance. Through 31 semi-structured interviews in China, we examined how participants judged whether wearable outputs were reliable enough for everyday use. Participants relied on brand and price, visible interface activity, lived interaction experience, and comparison with bodily sensation. These cues supported conditional trust, but did not reveal sensor validity, data continuity, or failure conditions. We describe this mismatch as an observability gap and outline design directions for showing signal quality, reliability by context, human-system fit, and alert provenance.
\end{abstract}



\maketitle

\section{Introduction}

An older adult deciding whether to rely on a health wearable rarely sees the properties that matter most from a systems perspective. They do not see sensor error rates, uncertainty bounds, data loss, model assumptions, or failure thresholds. They do see price, brand, medical-looking design, charts, update speed, and bodily comfort. Judgments of whether a system is ``good,'' ``accurate,'' or worth using are built from visible and embodied cues rather than direct access to system quality.

Wearable and app-based health monitoring now appear in self-tracking, remote care, and digital health infrastructures~\cite{guo2021review,hepburn2025,roos2023}. Reliable use depends on properties that are hard to inspect directly, including measurement validity, robustness, data continuity, wearability, interpretability, safety boundaries, and privacy~\cite{canali2022,carrier2020,fuller2020}. We call this problem an \emph{observability gap}. The term refers to a mismatch between the evidence users can inspect and the hidden properties they need in order to judge reliance.

Everyday reliance decisions do not wait for formal validation. A wearable may prompt a user to rest, ignore a warning, repeat a measurement, seek help, or keep using an uncomfortable but persuasive system. For older adults, the question is not only whether a device is adopted. It is how users decide what to believe after adoption, as the system keeps producing numbers, charts, and prompts.

Several HCI literatures explain nearby problems. Studies of sleep tracking show how users reason about credibility when sensing remains opaque~\cite{liang2016sleep,liang2020does}. Work on self-tracking app selection shows how people choose tools before long-term use begins~\cite{lee2021}. Research with older adults documents adoption, adherence, and abandonment around activity trackers~\cite{kononova2019,vargemidis2021}. These literatures explain credibility, selection, and uptake. They say less about how older adults infer whether a wearable-centered system is reliable enough for use once the sensing pipeline remains hidden.

This poster addresses that question through 31 semi-structured interviews with older adults who had at least four months of experience using health-related wearables. We report three preliminary insights. Older adults infer wearable quality under computational opacity. Visible evidence can diverge from hidden reliability properties. Interfaces can better show signal support, reliability by context, human-system fit, and alert provenance.

\section{Related Work}

\subsection{Reliance-Relevant Quality in Health Wearables}

Wearable-computing research defines quality through validity, reliability against reference devices, robustness outside controlled settings, and ecological validity~\cite{canali2022,carrier2020,fuller2020,roos2023}. Studies of person-generated wearable data highlight missingness, variable wear time, motion artifacts, and data fitness for downstream use~\cite{braem2024,cho2021a,vanderdonckt2024}. Long-term viability also depends on comfort, unobtrusiveness, privacy, and security~\cite{pettys2024,salehzadeh2024,stuart2024}. This work offers a researcher-centered account of system quality. It says less about how users approximate these qualities when they cannot inspect them directly.

\subsection{Adjacent HCI Work on Credibility, Selection, and Everyday Use}

Liang and Ploderer show that sleep-tracking users struggle to assess how devices produce their claims~\cite{liang2016sleep,liang2020does}. Our study broadens this question beyond sleep tracking. Lee et al. show that app selection is shaped by feature expectations, imagined future use, and social recommendation before long-term experience begins~\cite{lee2021,zeng2025parental}. Our study focuses on what happens after use becomes routine~\cite{meng2025tracing}. Work on older adults and activity trackers documents adoption barriers, motivators, and abandonment~\cite{gadahad2022,kononova2019,vargemidis2021}. We shift from adoption to inference, asking how lived experiences become judgments such as ``accurate,'' ``professional,'' or ``trustworthy.''

\subsection{Opacity, Explanations, and User Inference}

Research on explainability in wearable data analytics and personal sensing shows that explanations, personalization, and agency shape whether users accept system outputs~\cite{abdelaal2024,newn2022,he2023exploring,luo2025s}. Work on actionable sensing argues that detection alone is not enough because sensed data must be usable for real decisions~\cite{adler2024,meng2026tibetcpr}. Research on opaque algorithmic systems shows that users construct folk theories from partial cues and visible consequences~\cite{devito2018,eslami2016,meng2026living,liu2025supporting}. Wearable-centered health monitoring differs because bodily experience provides another source of evidence, and the stakes concern everyday self-management~\cite{meng202652}. We extend this work by showing how embodied calibration and conditional trust shape reliance judgments in health wearables. We use six quality domains from prior work as an analytical lens. Validity, context-specific reliability, data continuity, and human-system fit define the paper's core notion of reliance-relevant quality. The other two domains concern safety and alert provenance, and privacy and governance. We treat them as adjacent high-stakes qualities.

\section{Method}

\subsection{Study Design}
We conducted a qualitative interview study to examine how older adults make sense of health-related wearables in everyday life. We focused on the cues and experiences through which participants formed judgments about system capability and trustworthiness. We did not evaluate the technical performance of the devices they used. Terms such as ``accurate'' and ``reliable'' in the findings therefore denote participants' judgments rather than independently verified performance. We use \emph{wearable-centered health monitoring systems} to refer to body-worn sensing systems, companion applications, and home monitoring interfaces participants described as part of the same everyday monitoring practice. Only after participant-centered themes had been developed inductively did we compare them with system-quality dimensions from prior work.

\subsection{Participants and Recruitment}

We recruited 31 older adults in China, ages 61 to 76 years, $M = 68.4$. The sample included 16 men and 15 women, with 17 rural and 14 urban participants. Education included 16 participants with primary school or below, 8 with junior high school, 6 with high school, and 1 with a bachelor's degree. Reported health concerns included heart disease, obesity, hyperuricemia, hyperlipidemia, diabetes, hypertension, sleep disorder, and stroke. All participants had at least four months of experience using at least one wearable-centered health monitoring system. This threshold helped them report routine integration, breakdowns, switching decisions, and trust recalibration rather than first impressions. Participants described wrist-worn multi-sensor systems with companion apps, single-purpose home monitoring systems, and hybrid configurations. Recruitment used community notices and local networks.

\subsection{Data Collection and Analysis}

All interviews were conducted in person and lasted 45 to 63 minutes. The study received IRB approval from the University of Shanghai for Science and Technology. Participants provided written informed consent, were free to withdraw at any time, and received compensation of 50 RMB. Interviews were audio-recorded with consent, transcribed verbatim in Chinese, and translated into English with checks for conceptual fidelity. The semi-structured interview guide covered everyday use contexts, device histories, first encounters, and initial expectations. It also asked about memorable successes and breakdowns, switching, reduced or abandoned use, and comparisons across devices and with bodily sensations. Further questions examined how trust changed across situations and over time. We used concrete narrative prompts to elicit specific episodes rather than abstract evaluations.

We conducted reflexive thematic analysis~\cite{braun2006,braun2021one}. Analysis proceeded through familiarization and memo writing, line-by-line open coding, code comparison, provisional coding structure development, theme refinement across the full dataset, and comparison with system-quality dimensions from prior work. After about 24 interviews, additional data elaborated existing mechanisms rather than adding new ones. We continued to 31 interviews to test the stability and limits of the analysis. Consistent with reflexive thematic analysis, we did not report inter-rater reliability as a quality criterion~\cite{braun2021one}. Rigor was pursued through repeated returns to raw data, collaborative interpretation, translation checks, and attention to disconfirming cases.

\section{Findings}

Participants did not directly evaluate computational validity, data completeness, or uncertainty handling. Instead, they assembled judgments from everyday evidence. Across the dataset, four recurrent inference resources appeared. They were proxy cues of investment and reputation, visible signs of capability and autonomy, lived interaction experience, and experiential calibration against body and context. We call them \emph{recurrent resources} rather than stages. They often accumulated over time, but not every participant described all four. Earlier inferences were revised, reinforced, or reweighted as experience accumulated.

\subsection{Proxy Cues of Investment and Reputation}
Before participants had much direct experience, they relied on cues that were immediately visible and socially legible. Price, appearance, and brand were repeatedly used as signals of whether a device was ``serious'' enough to trust. Participants commonly assumed that a more expensive device must have involved more development effort or better components. Devices that looked medically styled or were associated with well-known brands were seen as more professional and trustworthy.

P6 described choosing by price: \textit{``I was choosing between two, one costing over seventy and one over three hundred. The interfaces looked similar, but I chose the more expensive one. Something that cheap can't really measure accurately, more like a toy. You get what you pay for.''} P13 made a similar inference from brand reputation: \textit{``I've seen similar devices from less well-known brands, but I don't dare use them. For big brands, even without knowing how they work, I feel they wouldn't cut corners.''} These were reputational rather than performance-based cues. They pointed to presumed seriousness, accountability, and development effort rather than to directly inspected sensing quality. Medical-looking appearance sometimes reinforced this logic by giving the device a clinical aura even in the absence of formal evidence.

Participants were not simply confusing price with accuracy. They used price, brand, and appearance because few stronger cues were available at selection. These cues made a device feel accountable before participants had enough experience to judge its outputs. They also persisted later as background assumptions that shaped how breakdowns were interpreted.

\subsection{Visible Signs of Capability and Autonomy}

Once participants began using a system, judgments shifted from what the system \emph{signaled} to what it appeared to do. Interfaces with curves, segmented charts, multiple indicators, or frequent updates were often taken as signs that the system was ``doing more'' internally. Proactive reminders and automated suggestions were interpreted as evidence that the system was actively sensing, analyzing, and thinking on the user's behalf.

This inference was clearest when participants compared sparse and dense interfaces. P2 said, \textit{``One I used before only had a heart rate number. It felt very 'empty,' like it wasn't doing anything. I even suspected the data might be made up. But this one is different. It has curves and time-segmented graphs. Although I can't really explain what they mean, I feel it is analyzing a lot of things.''} The issue was not comprehension alone. Dense output made invisible analysis feel present.

Update speed worked similarly. P21 said, \textit{``This one keeps updating. For example, if I just take a few steps, it changes immediately, so I feel it is constantly monitoring. But some I used before didn't change for a long time, and I would suspect they weren't calculating at all.''} Fast change made sensing feel temporally coupled to the body, while slow updates suggested that the system was not really working.

The denser the display, the faster the update, and the more agentic the behavior, the more participants treated the system as capable. ``AI'' labels often functioned as symbolic confirmation. P5 said, \textit{``It says AI sleep analysis. I don't really know how it works, but it feels more advanced than ordinary ones.''} Rather than inviting scrutiny, the label gave participants a culturally recognizable shorthand for intelligence. This did not distinguish genuine signal sensitivity from a polished feedback rhythm.

\subsection{Lived Interaction Experience}

With continued use, participants judged systems through the experience of living with them. Smoothness of use, bodily comfort, charging burden, and breakdowns became evidence of overall quality. If a system felt stable and easy to live with, it was often judged as a good system overall.

Participants treated small interaction failures as evidence about the whole system. P24 said, \textit{``There was one I used before for real-time blood pressure detection. After clicking into it, I had to wait for a while, and sometimes it would lag a bit, so I felt it wasn't very good, as if it wasn't very stable.''} P7 described repeated disconnections in similar terms: \textit{``Sometimes it disconnects and needs to be reconnected. If that happens many times, I feel it is not very stable, and I don't really dare to fully trust its data.''}

Bodily discomfort worked in the same evaluative register. P14 said, \textit{``There is one that feels a bit tight after wearing it for a long time, or is not very comfortable when sweating, so I don't really want to wear it anymore, and I also feel that this device is not made very well.''} Lag, disconnection, or discomfort were therefore not treated as minor peripheral issues. They were folded into broader judgments about whether the system itself was dependable.

A boundary case shows that participants were not simply maximizing comfort. P10 noted, \textit{``System performance is definitely more important, such as heavier weight and more sensitive sensors. This should sacrifice some comfort, and I don't think that is a very big problem.''} Participants balanced experiential fit with assumptions about what better sensing might require.

\subsection{Experiential Calibration Against Body and Context}

Over longer periods of use, participants developed practical rules for deciding when to trust the system by comparing outputs with bodily sensation and situation, rather than gaining direct access to internal sensing logic.

Participants often used bodily agreement as a practical check. P16 said, \textit{``Sometimes it says I slept badly, but I actually don't feel much at all, so I won't take it too seriously. But if I really feel very tired that day, and it also shows that I slept badly, then I feel it is right.''} The reading became more believable when the device and felt fatigue aligned.

P22 used exercise in the same way: \textit{``Once after exercising, the data was especially high, and I could clearly feel my heartbeat was very fast, so that time I felt it was quite accurate. But sometimes I don't feel much, yet it says it is very high, and then I feel it is a bit exaggerated.''} These checks also became context specific. P4 said, \textit{``I found that it was quite accurate when I was running, and the changes kept up pretty well. But at ordinary times, for example when sitting, sometimes the data was a bit strange, so I felt it was different in different situations.''}

Participants were not attempting formal validation. They built context-bound consistency rules. They trusted a reading more when it matched the body, less when it conflicted with felt experience, and only for some functions or situations. This was the most developed form of judgment in our data, but it still did not amount to transparent understanding.

\subsection{How Conditional Trust Accumulated}

These resources produced an accumulating pattern rather than a rigid sequence. Initial expectations from proxy cues often persisted. Visible capability and interaction experience could reinforce or weaken them. Embodied calibration sometimes overrode them in specific contexts. Longer-term use produced conditional trust grounded in visible cues, salient interactions, and embodied comparison, not transparent understanding.

Participants did not always choose the cue that looked most convenient or reassuring. A reputable device could lose credibility after repeated lag. A comfortable device could be treated as less powerful. A dense interface could remain attractive while its advice was downgraded after bodily mismatch. These cases show that conditional trust was not blind acceptance. It was a practical settlement among imperfect cues.

\section{Discussion}

\subsection{Users Infer What Systems Make Legible}

Participants inferred wearable quality from what systems made visible. The \emph{observability gap} names the mismatch between visible cues and the properties that matter for reliance. Current systems expose polish, activity, and symbolic sophistication more clearly than signal support, failure conditions, or data continuity. When computational properties were hidden, participants turned to price, brand, medical appearance, interface density, update frequency, automated prompts, and ``AI'' labels. These cues became substitutes for hidden qualities such as sensing validity and reliability. This finding aligns with HCI work on opaque systems, where people construct folk theories from visible cues and outcomes~\cite{devito2018,eslami2016}. It also complements work on personal sensing that shows how interface framing shapes acceptance of sensed outputs~\cite{abdelaal2024,newn2022}. The risk is direct. Visible sophistication can be mistaken for evidence of competence.

\subsection{Conditional Trust Is Not the Same as Understanding}

Longer-term use recalibrated trust without producing accurate understanding. Participants layered experience on top of earlier proxy judgments and developed local rules for when the system should be trusted. This is calibration, but not validation against a gold standard. In Lee and See's terms, the challenge is not whether people trust automation at all, but whether trust is appropriately calibrated to capability~\cite{lee2004trust}. Our participants were calibrating trust through embodied heuristics and local experience rather than through reliable access to signal quality, uncertainty, or failure conditions. That can be adaptive in daily life, yet is also fragile. If bodily sensations are delayed, ambiguous, or misleading, then ``it matches how I feel'' can reinforce error as easily as correct it. A system can be repeatedly used and trusted over time without users forming a technically accurate mental model of how it works.

\subsection{What Is Specifically Older-Adult About This Account}

In this older adult sample, several aspects were especially salient. Embodied calibration mattered because participants described long-term bodily routines, chronic-condition management, and familiar symptom patterns. The body served as a reference point for checking outputs. Comfort and routine fit also carried weight because participants evaluated systems as things to live with over time, not as short-term novelties. Discomfort undermined the plausibility of the whole system as a sustainable aid~\cite{keogh2020,ding2024,stuart2024}. The stakes of overtrust and mistrust may also be higher when wearable outputs enter self-management or telehealth routines~\cite{hepburn2025,wang2025,chen2026between}.

\subsection{Design Implications for Reliability Evidence}

The design challenge is to show the evidence users need at the moment they judge reliance. The goal is not to add information or make systems feel intelligent. We outline four directions that address signal quality, context-specific reliability, human-system fit, and alert provenance.

This suggests a design criterion. Reliability evidence should appear where users already make judgments, not only in help pages, manuals, or privacy settings. Participants judged systems while reading a graph, responding to an alert, noticing lag, charging a device, or comparing an output with how they felt. These moments are where interfaces can show uncertainty, signal support, and limits without requiring technical documentation.

\textbf{Surface signal quality and sensing gaps.} Participants treated continuous updates as evidence that a system was working correctly. Trustworthy outputs also depend on signal quality, wear adherence, and continuity~\cite{braem2024,cho2021a,cho2021b,vanderdonckt2024}. Interfaces should show how well an estimate is supported. Contact quality indicators, wear time summaries, and low-confidence notices would give users better grounds for deciding when to rely on an output.

\textbf{Communicate reliability by context rather than universal authority.} Participants already built local trust rules through trial and error. Systems should not leave this entirely to user guesswork. Rather than blanket claims such as ``accurate monitoring,'' designers could specify when outputs are most and least reliable. This direction is consistent with work showing that personalization and deployment context shape sensing performance~\cite{han2024,li2024,meegahapola2023generalization}.

\textbf{Treat human-system fit as part of reliance.} Comfort, charging burden, and friction were folded into global judgments of whether the system was well made. Designers should treat wearability as part of reliance. A device users cannot comfortably sustain will struggle to produce the continuity needed for meaningful monitoring~\cite{keogh2020,ding2024,stuart2024}.

\textbf{Distinguish automation from medical authority.} Participants often interpreted automated prompts as signs that the system was ``thinking for them,'' yet automation can signal convenience without guaranteeing clinical validity. Systems should distinguish between wellness prompts, heuristic suggestions, and clinically grounded alerts. Labeling the provenance and status of each alert is one practical way to keep users appropriately in the loop~\cite{adler2024,arakawa2023prism,gathright2024,kianpisheh2024exhar}.

\subsection{Limitations}
This study recruited older adults in China who used consumer or hybrid wearable systems. The cues participants used and their relative salience may reflect the local wearable market and care context. We do not assume that these patterns transfer unchanged to older adults in other settings. The data come from retrospective interviews rather than direct observation. We did not independently validate the devices participants used. The contribution is not a technical assessment of wearable accuracy. It is an exploratory account of how available cues do not map cleanly onto reliance-relevant qualities. Privacy and governance rarely surfaced spontaneously, so we treat them as future work rather than a primary empirical claim. These design directions have not been evaluated. Future work should translate them into prototypes for wearable systems used by older adults. Such studies should examine whether the prototypes support appropriately calibrated reliance without increasing interpretation burden. Future work should also combine interviews with diaries, observation, or logs to compare reported trust changes with breakdowns, missing data, and context shifts.


\end{document}